\documentclass[12pt]{article}
\usepackage{epsfig}
\begin{document}
\title{Force Relaxation in the q--model for Granular Media} 
\author{ Jacco H. Snoeijer and J. M. J. van Leeuwen\\
Instituut--Lorentz, Leiden University, \\
P.O. Box  9506, 2300 RA Leiden, The Netherlands}
\date{\today} 
\maketitle

\begin{abstract}
We study the relaxation of force distributions in the q-model, assuming
a uniform q-distribution. We show that "diffusion of correlations" makes
this relaxation very slow. On a $d$--dimensional lattice, the asymptotic
state is approached as $l^{(1-d)/2}$, where $l$ is the number of layers
from the top. Furthermore, we derive asymptotic modes of decay, along
which an arbitrary short--range correlated initial distribution will
decay towards the stationary state. 
\vspace*{3mm}

\end{abstract}

\section{Introduction} \label{intro}
To answer basic questions with simple models is the favorite topic in the work of 
Bob Dorfman. This contribution, dedicated to him on the occasion of his 65th 
anniversary, concerns the force relaxation in a model for granular media.
The model is certainly simple but whether the question we answer is a basic one,
we leave to his judgment.

\begin{figure}[h]
   \centering{ \epsfxsize=7cm
   \epsffile{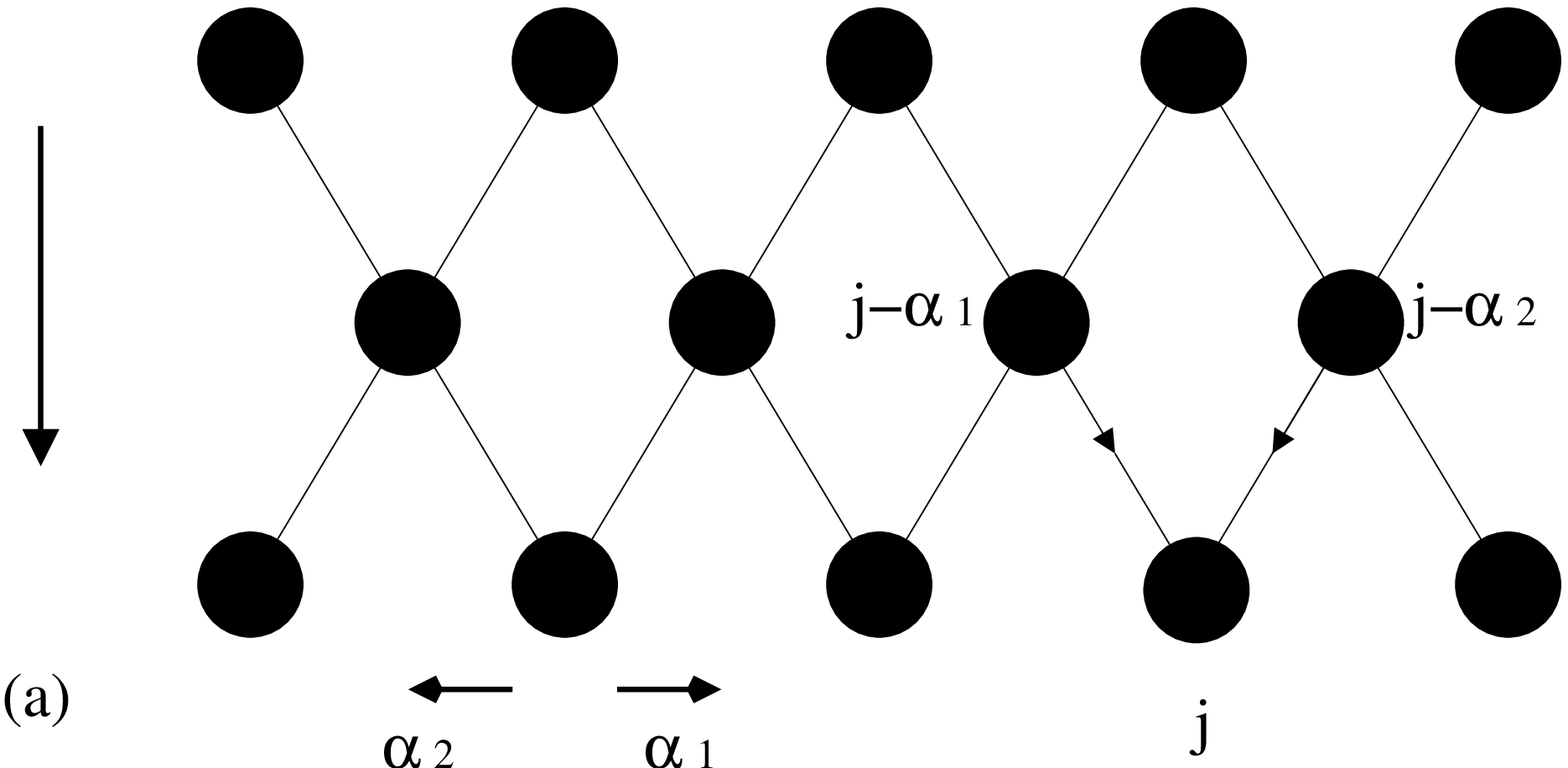} \epsfxsize=7cm \epsffile{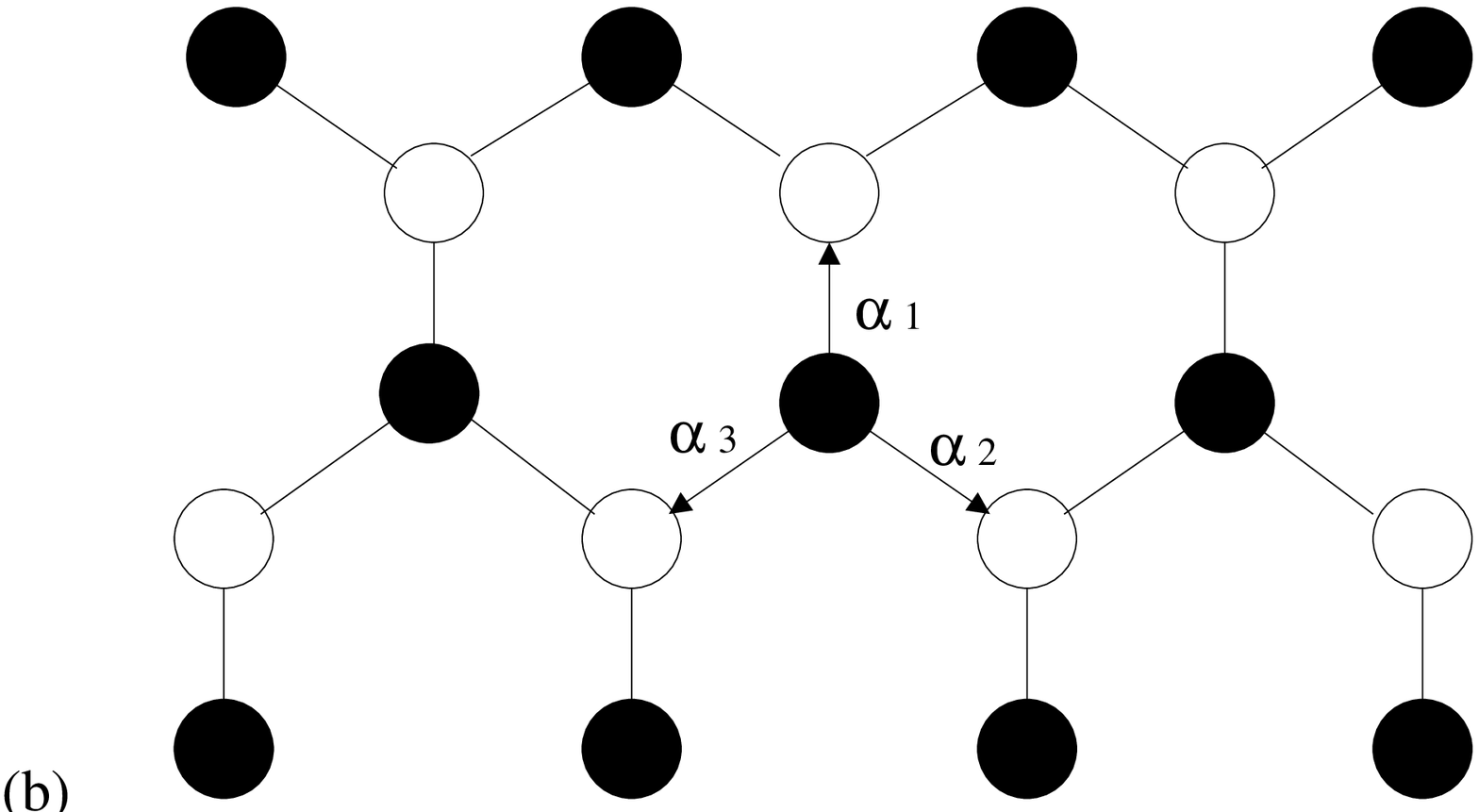}}
   \caption{The displacement vectors $\alpha$ in the q--model for (a) the triangular packing 
(side view) and (b) the fcc packing (top view) }
   \label{qmd}
\end{figure}

The first idealization of granular media is to consider them as a pack of equally sized 
beads. The second idealization is to put the beads on a regular lattice, with periodic
boundary conditions horizontally in order to avoid boundary effects. The problem is
the propagation of the force exerted on the top layer downwards to the deep lower 
layers of the bead pack. Now it might seem that all randomness, characteristic for 
granular media, has been lost in this idealization. However the forces which the
beads exert on each other are supposed to be transmitted in a random way. Let $f_i$ be 
the force in the downward direction on the $i$th bead in a layer. 
This bead makes contact with a number of $z$ beads in the layer below, 
which we indicate by the indices $i+\alpha$. The $\alpha$'s  are displacement 
vectors in the lower layer as shown in Fig. \ref{qmd} (a) for a 2--dimensional triangular 
packing and in (b) for a 3--dimensional fcc packing.
Bead $i$ transmits a fraction $q^\alpha_i$ of the force $f_i$ to the bead 
$i+\alpha$ underneath it. The random element in the model is that 
the fractions $q^\alpha_i$ are taken stochastically from a uniform distribution satisfying 
the constraint
\begin{equation} \label{a1}
\sum_\alpha q^\alpha_i = 1.
\end{equation}
This is the q--model as introduced by Liu et al. \cite{liu}. The problem we consider is:
given the force distribution $P_0 (f_1, \cdots, f_N)$ in the top layer, what 
is the force distribution in the lower layers and in particular how does the
distribution approach its asymptotic value? Recent studies \cite{raj,lew} addressed
the relaxation of the second moments for general $q$ distributions on a triangular lattice.
Restricting ourselves to the uniform $q$ distribution, we investigate the relaxation of the
full distribution for arbitrary lattices.

In order to write the equation for
the force distribution we consider the force $f'_j$ on the $j$th bead in 
layer. It is composed of forces $f_{j-\alpha}$ from the layer on top of it as
\begin{equation} \label{a2}
f'_j = \sum_\alpha q^\alpha_{j-\alpha} f_{j-\alpha}.
\end{equation}
We have left out the so-called injection term representing the weight of the particles.\footnote{
Leaving out the injection term is legitimate when the applied force is large compared to the
particle weight or when the direction of propagation is perpendicular to gravity.}
As a consequence of the regularity of the lattice, a bead is supported by 
$z$ beads in the next layer and reciprocally a bead in the lower layer also has $z$ beads 
pressing on it. Note that, due to the fluctuations in the $q$'s, the sum over the 
$q^\alpha_{j-\alpha}$ in (\ref{a2}) does not  add up to 1, which means that even  a 
set of equal $f_{j-\alpha}$ will lead to fluctuations in the $f'_j$.

The transformation of the force distribution from one layer to the next layer
can be written as
\begin{equation} \label{a3}
P' (\vec{f}\,') = \int d \vec{f} \int D \vec{q} \prod_j \delta (f'_j - 
\sum_\alpha q^\alpha_{j-\alpha} f_{j-\alpha} ) \,P (\vec{f}),
\end{equation}
where we have introduced a vector notation for the forces in one layer 
$\vec{f} = (f_1, \cdots , f_N)$, and for the integrations we use the abbreviations
\begin{equation} \label{a4}
\int d \vec{f} = \prod_i \int^\infty_0 df_i \, , \quad \quad \quad \int D \vec{q} = 
\prod_i \, D q_i = \prod_i \, (z-1)! \,\prod_\alpha \int^1_0 d q^\alpha_i \,\delta 
\left( \sum_{\alpha'} q^{\alpha'}_i -1 \right).
\end{equation}
As one sees from the integration over the $q^\alpha_i$, we have taken a uniform and
normalized distribution over these stochastic variables. The restriction to a 
uniform $q$--distribution simplifies the mathematics while it still reproduces 
important experimental observations \cite{exp}.

The recursive relation (\ref{a3}) between two successive layers contains the
problem of force relaxation from top to bottom of the bead pack. For a number of properties
it is more convenient to work with the Laplace transform of the distribution,
\begin{equation} \label{a5}
\tilde{P} (\vec{s}) = \int d \vec{f} \exp( -\vec{s} \cdot \vec{f}) \, P (\vec{f}).
\end{equation}
The Laplace transform of (\ref{a3}) leads to
\begin{equation}\label{a6}
\tilde{P}' (\vec{s}) = \int D \vec{q} \int d \vec{f} \exp \left(-\sum_j \sum_\alpha s_j 
q^\alpha_{j-\alpha} f_{j-\alpha} \right) \, P (\vec{f}).
\end{equation}
The sum in the exponent can be rewritten as
\begin{equation} \label{a7}
\sum_j \sum_\alpha q^\alpha_{j-\alpha} \,s_j \,f_{j-\alpha} =
\sum_i \left( \sum_\alpha q^\alpha_i \,s_{i+\alpha} \right) f_i,
\end{equation}
by changing the summation variable $j$ to $i = j-\alpha$. So we can express
the right hand side of (\ref{a6}) also in the Laplace transform and we get
\begin{equation} \label{a8}
\tilde{P}' (\vec{s}) = \int D \vec{q} \, \tilde{P} (\vec{s} (\vec{q})),
\end{equation}
with
\begin{equation} \label{a9}
s_i (\vec{q}) = \sum_\alpha q^\alpha_i \,s_{i+\alpha}. 
\end{equation}
The projection of the full distribution $P (\vec{f})$ to the single--site force
distribution $p (f_i)$ is particularly simple in the Laplace language
\begin{equation} \label{ins1}
\tilde{p} (s_i) = \tilde{P} (0, \cdots, 0, s_i, 0, \cdots, 0),
\end{equation}
since $s_l=0$ means a full integration over $f_l$.

Either (\ref{a3}) or (\ref{a8}) completely defines the propagation of the
forces for boundary conditions $(N,M)$, where
$N$ is the number of sites in a layer and $M$ the number of layers, (both assumed to
be large). For the ultimate asymptotic behavior one has $M \gg N$, 
but the $N \gg M$ case is physically more relevant and therefore the main focus of this
paper. It is also not sensitive to the (periodic) boundary conditions that we have chosen.

Coppersmith et al. \cite{cop} derived a stationary state of (\ref{a3}),
\begin{equation} \label{a10}
P^* (\vec{f}) = \prod_i p^* (f_i)\, , \quad\quad \quad 
p^* (f) = z^z f^{z-1} \exp (-zf) /(z-1)!.
\end{equation}
It is a product state without any correlation between sites. They
provided numerical evidence that it is indeed the asymptotic force distribution.
This paper concerns the decay towards the stationary state, 
which turns out to be algebraic rather than exponential, as is implicit in \cite{cop}.
We will show how ``diffusion of correlations'' accounts for this slow relaxation. 
Furthermore, we will pay attention to the question of which initial 
distributions evolve to the stationary state (\ref{a10}) and in what sense the approach
has to be understood. 

The first part of this paper deals with the decay of the distribution on the level of its 
moments. We start out by showing that the distribution of the total force in a layer
is invariant under the recursion. Then we illustrate the propagation of forces 
in the system by considering the evolution of the first few moments of the force 
distribution.  In the second part of this paper we construct solutions of the recursion
relations, which dominate the asymptotic decay towards the stationary state (\ref{a10}), 
with an emphasis on the relaxation of the single--site force distribution.
In the concluding remarks we comment on the results that we have obtained and discuss
which of these can be generalized to arbitrary $q$ distributions.

\section{The distribution of total force} \label{total}

There is one obvious invariant in the problem: the total force on the particles
of one layer
\begin{equation} \label{x1}
F = \sum_i f_i.
\end{equation}
This can be seen by summing (\ref{a2}) over all $j$
\begin{equation} \label{x2}
F' = \sum_j f'_j = \sum_{j,\alpha} q^\alpha_{j-\alpha} f_{j-\alpha} = F,
\end{equation} 
by again changing the summation variable $j$ to $i=j-\alpha$. The magnitude of $F$ is 
irrelevant for the problem since, due to the 
linearity of the force law (\ref{a2}), an overall scaling of the forces is possible 
without changing the physics of the problem. Thus we could fix the value of $F$ and only
consider distributions having strictly this value, but that is mathematically not 
very convenient. The distribution of the
total force is obtained from the distribution of the forces as
\begin{equation} \label{x3}
R (F) = \int d \vec{f} \,\delta (F - \sum_i f_i) \,P (\vec{f}).
\end{equation}
Since  relation (\ref{x2}) holds for any set of $q$'s, one finds from (\ref{a3}) that 
$R(F)$ is invariant.

So the initial distribution $R^{(0)} (F)$ dictates that of the asymptotic state. 
This seems a strong restriction on the relaxation of the distribution function, 
but in practice it is rather a warning on what quantities to inspect. As an example 
consider the total force distribution of the stationary state (\ref{a10}), which reads
\begin{equation} \label{x6}
R^* (F) = z^N \, F^{zN-1} \, \exp (-z F) / (zN-1)!.
\end{equation}
It is a sharp distribution with a mean and variance
\begin{equation} \label{x7}
\langle \, F \, \rangle = N, \quad \quad \quad \langle \, 
F^2 \, \rangle -\langle \, F \, \rangle^2 = N/z.
\end{equation}
Note that fixing the total force $F$ excludes (\ref{a10}) as the stationary state!
The distribution (\ref{x6}) is very reminiscent of the energy distribution 
in the canonical  ensemble. Indeed this analogy, also mentioned in \cite{cop},
is illuminating. For large $N$ 
the total force distribution (\ref{x6}) is narrow, in the same way as the total energy
distribution in the canonical ensemble.
Now it is well known that the correlation functions of the canonical ensemble and the
micro--canonical ensemble (in which the energy is fixed), 
coincide for all distances (to order $1/N$). But integrals over the correlation functions
over all distances differ in the two ensembles. A similar subtlety arises here. As long
as we consider spatial correlations between forces, the distribution of the total 
force is unimportant, if it is sharp in the sense of (\ref{x7}). However, for sums 
over the correlations, differences will appear. Consider for instance the force
correlations between two sites, defined as
\begin{equation} \label{new1}
\langle \, f_i f_{i+n} \, \rangle = \int d\vec{f}\,  f_i f_{i+n} \,P (\vec{f}). 
\end{equation}
Summing $\langle f_i f_{i+n} \rangle$ over $i$ and $n$ yields
\begin{equation} \label{new2}
\sum_{i,n} \langle \, f_i f_{i+n} \, \rangle = \int d\vec{f}\, \sum_{i,n}
f_i f_{i+n} \,P (\vec{f})= \langle \, F^2 \, \rangle. 
\end{equation}
If $F$ is allowed to fluctuate, this sum clearly differs from the result obtained for 
fixed values of $F$. The fluctuations which we allow in the total force are limited to
\begin{equation} \label{vn1}
\langle \, F^2 \, \rangle -\langle \, F \, \rangle^2 = c N,
\end{equation}
with $c$ of order 1. Then they are unimportant in the thermodynamical limit.

\section{Diffusion of the first moment} \label{moment}

Forces applied to one bead in the top layer will diffuse into a wider region in the
lower layers. To see this, consider the  first moments $\langle \, f_j \, \rangle $
of the distribution for an inhomogeneous initial distribution. 
The relaxation of these moments is given by
\begin{equation} \label{y6}
\langle \, f_j \, \rangle' =
\int d \vec{f} \int D \vec{q} \,\sum_\alpha q^\alpha_{j-\alpha} \, f_{j-\alpha} \,P(\vec{f}) =
\int D \vec{q} \,\sum_\alpha q^\alpha_{j-\alpha} \langle \, f_{j-\alpha} \, \rangle.
\end{equation}
Integrals over the $q'$'s are elementary, but this one is trivial 
as each $ q^\alpha_{j-\alpha}$ gives the same answer, and with (\ref{a1}) it must be
equal to $1/z$. So the result becomes
\begin{equation} \label{y7}
\langle \, f_j \, \rangle' = {1 \over z} \sum_\alpha 
\langle \, f_{j-\alpha} \, \rangle.
\end{equation}
This simple recursion law tells us that a distribution with only a non--vanishing
moment on one site in the top layer gradually spreads over the lower layers like a 
diffusion process. 

In order to understand the role of system size it is useful to inspect the fourier transform
\begin{equation} \label{y8}
m ({\bf k}) = \sum_i \langle \, f_i \, \rangle \exp (i {\bf k \cdot r}_i),
\end{equation}
which relaxes from layer to layer as
\begin{equation} \label{y9}
m' ({\bf k}) = \lambda ({\bf k}) \,m ({\bf k}),
\end{equation}
with the transmission function
\begin{equation} \label{y10}
\lambda ({\bf k}) = {1 \over z} \sum_\alpha \exp (i {\bf k \cdot r}_\alpha).
\end{equation}

In accordance with the conservation of the total force we see that the mode
${\bf k=0}$ is conserved as $\lambda ({\bf 0}) = 1$. All the other modes 
decay exponentially since $|\lambda ({\bf k \neq 0})| < 1$.
So we find asymptotically
\begin{equation} \label{y12}
m^{(\infty)} ({\bf k}) = \delta_{\bf k, 0} \, m ({\bf 0}),
\end{equation}
with $m ({\bf 0})$ the average of the (initial) total force. Translating it back to space yields
\begin{equation} \label{y13}
\langle \, f_i \, \rangle^{(\infty)} = \langle \, F \, \rangle /N,
\end{equation}
i.e. all sites feel the same average force regardless of the initial distribution.

Clearly this result holds in the limit $M \gg N$. In the other limit $N \gg M$ we may
replace the summation over ${\bf k}$ by an integration over the Brillouin Zone (BZ) of
the layer and one gets the standard diffusion process from the integral
\begin{equation} \label{y14}
\langle \, f_j \, \rangle^{(M)} = {1 \over V_{BZ}} \int_{BZ} d {\bf k} \, m ({\bf k}) \,
\exp [\,M \log (\lambda ({\bf k})) - i {\bf k \cdot r}_j\,].
\end{equation}
In the limit of large $M$ the integral can be evaluated by the saddle point method,
using the fact that the dominant contribution comes from small ${\bf k}$. This allows
us to approximate $\lambda ({\bf k})$ by
\begin{equation} \label{y15}
\log [\,\lambda ({\bf k})\,] \simeq \log [\,1 - \sum_\alpha ({\bf k \cdot r}_\alpha)^2/2\,]
\simeq -D\, k^2,
\end{equation}
which renders the integration effectively a Gaussian. The diffusion constant $D$ depends 
on the latttice structure of the layer, with $D=1/8$
for the triangular packing and $D= 1/12$ for the fcc packing. 
Thus one sees that the ratio of $M/N$ decides whether we should take a discrete
sum over the ${\bf k}$ variables ($M \gg N$) or whether we should integrate over
the Brillouin Zone ($N \gg M)$. In the first case, only ${\bf k=0}$ matters and we 
have diffusion crossing the periodic boundary conditions in the layer. In 
second case, a region around ${\bf k = 0}$ has an influence and there are no effects 
from the finite size of the layer.

Having seen the washing out of spatial inhomogeneities by diffusion, we now concentrate 
on translational invariant initial conditions and on the case $N \gg M$.

\section{Evolution of the second moment} \label{corr}

As we are free to put $\langle \, f_i \, \rangle = 1$, the relaxational properties show up
only in the higher moments of the force distribution. We therefore examine 
how these moments evolve under the recursion. The transformation of second moments 
$\langle \, f_i f_{i+n} \, \rangle$ is obtained by combining (\ref{new1}) and 
(\ref{a3}) as
\begin{equation} \label{z2}
\langle \, f_j f_{j+n} \, \rangle' = \sum_{\alpha,\alpha'} \left( \int D \vec{q} \,
q^\alpha_{j-\alpha} \, q^{\alpha'}_{j+n-\alpha'} \right) \,\langle \, f_{j-\alpha} 
f_{j+n-\alpha'} \, \rangle.
\end{equation}
The $q$-integrals can be worked out and yield
\begin{equation} \label{z3}
\int D \vec{q}  \, q^\alpha_{j-\alpha} \, q^{\alpha'}_{j+n-\alpha'} =
{1 \over z^2} + \left(-{1 \over z^2 (z+1)} +
{\delta_{\alpha,\alpha'} \over z(z+1)} \right) \delta_{0,n+ \alpha -\alpha'}. 
\end{equation}
Inserting this into (\ref{z2}) gives the following recursion scheme
\begin{equation} \label{z4}
\langle \, f_j f_{j+n} \, \rangle' =  {1 \over z^2} \sum_{\alpha,\alpha'}  
\langle \, f_{j-\alpha} f_{j+n-\alpha'} \, \rangle + 
g_n \langle \, f^2_j \, \rangle. 
\end{equation}
The function $g_n$ incorporates all corrections due to the Kronecker $\delta$'s in 
(\ref{z3}) and the only non--zero terms are
\begin{equation} \label{z5}
g_0 = {z-1 \over z(z+1)}, \quad \quad \quad  g_\gamma = - {1 \over z^2 (z+1)}.
\end{equation}
The index $\gamma$ denotes a nearest neighbor position.
It is easily checked that the difference of two unequal displacement vectors $\alpha$ 
and $\alpha'$ points to a nearest neigbor position. It is a matter of counting to 
verify that correlations of the type
\begin{equation} \label{z6}
\langle \, f_i f_{i+n} \, \rangle^* = C (1 + {1 \over z} \delta_{0,n}) 
\end{equation}
are fixed points of relation (\ref{z4}) for arbitrary $C$. This conclusion was also derived 
by Lewandowska et al. \cite{lew} for the specific case of a 2-dimensional triangular 
lattice and $C=1$, (corresponding to (\ref{a10})). For $C \neq 1$ one has long--ranged 
correlations since, for large $n$, the average of the product does not approach the product 
of the averages (set equal to 1). $C$ is related to the scale of the 
forces as one sees from (\ref{new2}). This relation implies for (\ref{z6})
\begin{equation} \label{new3}
\langle \, F^2 \, \rangle^* = C \, (N^2 + N/z).
\end{equation}
For a sharp distribution of the total force in the sense of (\ref{x7}) and normalization
of the average $\langle F \rangle = N$, there is little room for $C$ to differ from 1,
allowing only deviations of order $1/N$. So we take $C=1$ corresponding to the stationary
state (\ref{a10}) and come back to the issue in the Appendix.

\begin{figure}[h]
   \centering \epsfxsize=\linewidth \epsffile{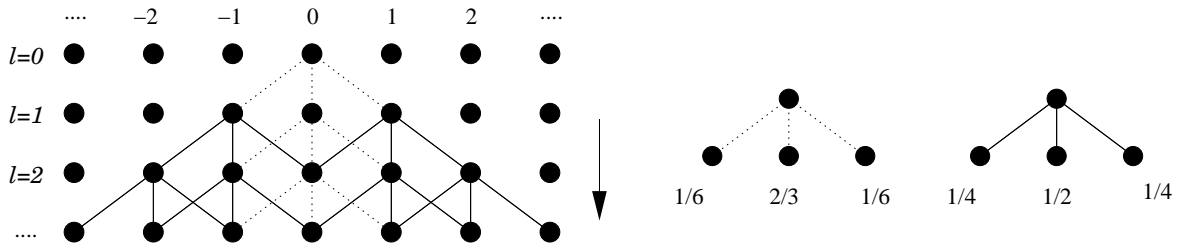} 
   \caption{The flow diagram for $A^{(l)}_n$ in the triangular lattice, starting with only
$A^{(0)}_0 \neq 0$. The horizontal index $n$ indicates the relative distance.}
   \label{flow}
\end{figure}

To study the relaxation in more detail we consider the
difference with respect to values of the stationary distribution 
\begin{equation} \label{z7}
A_n = \langle \, f_i f_{i+n} \, \rangle - \langle \, f_i f_{i+n} \, \rangle^*,
\end{equation}
which will of course relax according to the same relation (\ref{z4}). As an example,
the flow diagram for the $A_n$ is depicted in Fig. \ref{flow} for the triangular lattice.
The advantage of taking the difference is that its fourier 
transform is well behaved for distributions which have only short range correlations, 
since both the initial correlation function and the stationary values
approach 1 for large distances $n$. The fourier transforms are defined as
\begin{equation} \label{z8}
A ({\bf k}) = \sum_n A_n \exp (i {\bf k \cdot r}_n), \quad \quad \quad 
A_n = {1 \over V_{BZ}} \int_{BZ} d{\bf k} \, A ({\bf k}) \exp (-i{\bf k \cdot r}_n ),
\end{equation}
and $A ({\bf k})$ transforms as
\begin{equation} \label{z9}
A' ({\bf k}) =  \lambda ({\bf k}) \,\lambda (-{\bf k})\, A ({\bf k}) + A_0 \, g ({\bf k}).  
\end{equation} 
The correction function $g ({\bf k})$ can be written as 
\begin{equation} \label{z10}
g ({\bf k}) = g_0 + \sum_\gamma g_\gamma \,\exp (i {\bf k \cdot r}_\gamma) =
{1 \over z+1} [\, 1 - \lambda ({\bf k}) \,\lambda (-{\bf k}) \,].
\end{equation}
Note that $A ({\bf 0})$ is invariant, which implies
\begin{equation} \label{z11}
\sum_n A'_n = \sum_n A_n,
\end{equation}
as it should be according to relation (\ref{new2}).

The recursion (\ref{z9}) can be solved by introducing the generating function
\begin{equation} \label{z12}
A (u,{\bf k}) = \sum_{l=0} A^{(l)} ({\bf k}) \,u^l,
\end{equation}
where the sum extends over all layers $l$. Multiplying (\ref{z9}) by $u^{l+1}$ and
summing over $l$ gives an algebraic equation for the generating function
\begin{equation} \label{z13}
A (u,{\bf k}) - A^{(0)} ({\bf k}) = u \left[ \,\lambda ({\bf k}) \,\lambda (-{\bf k}) \,
A (u,{\bf k}) + A_0 (u) \,g ({\bf k}) \, \right],
\end{equation}
where the auxiliary function $A_0 (u)$ is defined as
\begin{equation} \label{z14}
A_0 (u) = \sum_{l=0} A^{(l)}_0 \,u^l.
\end{equation}
$A_0 (u)$ is our prime interest as it contains the center coefficients $A^{(l)}_0$, 
giving information about the relaxation of the single--site force distribution. 
The other $A^{(l)}_n$ contain correlations between two sites.
Using (\ref{z10}), we write the solution of (\ref{z13}) as
\begin{equation} \label{z15}
A (u,k) = \frac{A^{(0)} ({\bf k}) }{1 - u \, \lambda ({\bf k}) \,\lambda (-{\bf k})}
+ \frac{1}{z+1} A_0 (u)
\left(1 - \frac{1-u}{1 - u \,\lambda ({\bf k}) \,\lambda (-{\bf k}) } \right).
\end{equation}
The first term will be shown to give purely diffusive behaviour. This can be understood 
from the recursion of (\ref{z4}): without the corrections $g_n$, it becomes a regular 
random walk in a layer of dimension $d-1$. The second term, which originates from the 
$g_n$, contains a correction to the asymptotic amplitude and a term (proportional to
$1-u$) representing a transient. Hence, we anticipate diffusive relaxation 
$\sim l^{(1-d)/2}$.

Using Eq. (\ref{z8}), the function $A_0 (u)$ satisfies 
\begin{equation} \label{z16}
A_0 (u) = {1 \over V_{BZ}} \int_{BZ} d {\bf k} \, A (u,{\bf k}). 
\end{equation}
Inserting expression (\ref{z15}) into (\ref{z16}) gives a closed equation for
$A_0 (u)$.

As we shall see, the large $l$ behaviour is dominated by the behavior at small ${\bf k}$.
We have assumed the initial correlation to be short--ranged such that $ A^{(0)} ({\bf k})$
is a regular function of ${\bf k}$ at the origin. Therefore we may as well
continue by considering first the {\it uncorrelated} initial distributions. 
Then only the value $n=0$ gives a contribution and we have initially
\begin{equation} \label{z17}
A^{(0)} ({\bf k}) =  A^{(0)}_0, 
\end{equation}
as depicted in Fig. \ref{flow}.
$A^{(0)}_0$ sets the overal amplitude of the relaxation process. 

Before discussing the asymptotics of the general case we begin by giving the
results for the triangular packing with $z=2$. In this case all the integrals 
can be carried out explicitly by writing them as contour integrals in the complex 
$v =\exp (ik)$ plane. For (\ref{z16}) and (\ref{z15}) we get
\begin{equation} \label{z18}
A_0 (u)=\frac{A^{(0)}_0}{\sqrt{1-u}} + \frac{1}{3} A_0(u)\,
\left(1-\frac{1-u}{\sqrt{1-u}}\right).  
\end{equation}
This relation for $A_0 (u)$ can readily be solved as
\begin{equation} \label{z19}
A_0 (u)=\frac{A^{(0)}_0}{\sqrt{1-u}}\, \left(\frac{3}{2+\sqrt{1-u}} \right) = 
\frac{3 A^{(0)}_0}{2\sqrt{1-u}}\, \left(1-\frac{1}{2}\sqrt{1-u}+\frac{1}
{4}(1-u)- \cdots \right).
\end{equation}
We have made the expansion in powers of $\sqrt{1-u}$ because each higher term leads
to a weaker singularity and therefore to a faster decay. Tracing back these features to
the form (\ref{z15}), we indeed see that the first term is responsible for 
the leading singularity, that the second modifies the amplitude by a factor 3/2 and that
the third leads to higher powers and thus to transient behavior.
The leading singularity gives an expansion
\begin{equation} \label{z20}
\frac{1}{\sqrt{1-u}}=\sum_{m=0}^{\infty} \frac{1}{2^{2m}}\, \frac{(2m)!}{(m!)^2}\, u^m.
\end{equation}
Note that these are precisely the binomial coefficients one expects from the random walk
described by the first term in (\ref{z4}).
Using Stirlings formula for the factorials yields as the leading term for the $A^l_0$ 
\begin{equation} \label{z21}
A^l_0 \simeq  \frac{3 A^{(0)}_0}{2 \sqrt{\pi l}}, \quad \quad \quad l \rightarrow \infty.
\end{equation}

So we see that the singularity at $u=1$ contains the asymptotic 
behaviour. This is also true for the general case. Defining the integral
\begin{equation} \label{z22}
D (u) = {1 \over V_{BZ}} \int_{BZ} d {\bf k} \, {1 \over 1 - u \, \lambda ({\bf k}) 
\lambda (-{\bf k})},
\end{equation} 
we get for $A_0 (u) $
\begin{equation} \label{z23}
A_0 (u) = A^{(0)}_0 \,D (u) + {1 \over z+1} A_0 (u) \, [\, 1 - (1-u) D (u)\,].
\end{equation}
We again have not written the solution in closed form in order to see the origin of the
terms. The first term contains the diffusive singularity at $u=1$. This follows from
the integral for $D(u)$ which develops a singular denominator for $u=1$ at the
point ${\bf k=0}$. The type of singularity depends on the dimension $d-1$ of the ${\bf k}$
integral and thus on the dimension of the system. The singularity can be obtained by the 
saddle point method using the approximation (\ref{y15}) for $\lambda ({\bf k})$ and reads
$(1-u)^{(d-3)/2}$, where the power 0 means a logarithm (for the fcc packing). 
Consequently the decay is as $l^{(1-d)/2}$ for large $l$. The other terms in (\ref{z23})
give a modification of the amplitude by a factor $(z+1)/z$ and a faster decay term 
(due to a weaker singularity).

In setting (\ref{z17}) we have made the restriction to initial states in which the second 
moments of the forces are uncorrelated.
From the analysis it is clear that for the asymptotics it is a justified substitution
for any initial state with short--ranged correlations. Thus the conclusion of this section
is that for arbitrary initial distributions with short--ranged initial correlations the second 
moment of the force  correlations relax towards the stationary distribution (\ref{a10}).
To be more precise, the approach is not perfect in view of  the restriction (\ref{z11}). 
The values of all $A^{(l)}_0$ saturate at a level $\sim A^{(0)}_0/N$ 
when $l$ starts to exceed the value $N^{2/(d-1)}$. 
Then the diffusion of the initial deviation goes around the periodic boundary conditions
and the system will approach a state (\ref{z6}) with $C-1$ of the order $1/N$.
This is the subtlety to which we alluded in Section \ref{total}. The total sum over
the $A_n$ corresponds to 
\begin{equation} \label{z24}
\sum_n A_n = (\langle F^2 \rangle - \langle F^2 \rangle^*)/N,
\end{equation}
and when $R^{(0)}(F)$ is not perfectly equal to $R^*(F)$, the system will never 
fully relax to $P^*(\vec{f})$.

\section{Stationary State and some Properties} \label{prop}

Coppersmith et al. \cite{cop} have determined a stationary state of Eq.  (\ref{a8}),
which is special in the sense that it is a product of single--site distributions.
Since the proof of the stationary distribution is vital for the decay
of deviations, we briefly repeat its construction. It is based on the following identity 
which holds for uniform $q$ distributions \cite{zinn}
\begin{equation} \label{b1}
\int D q_i \, \left(\frac{1}{1 + \lambda \sum_\alpha q^\alpha_i b_\alpha}\right)^z 
=\prod_\alpha \,\frac{1}{1+ \lambda b_\alpha}. 
\end{equation}
Now if we apply this identity to the (normalised) distribution
\begin{equation} \label{b2}
\tilde{P} (\vec{s}) = \prod_i \left( \frac{1}{1 + \lambda s_i} \right)^z,
\end{equation}
we see that each $z$th power of the factor splits up into $z$ single powers 
of the neighboring sites. The product over all sites in a layer then restores the
$z$th power of the denominator, which makes $\tilde{P} (\vec{s})$ an invariant. 
Putting the mean force equal to 1 requires that $\lambda =1/z$. So we have a stationary 
distribution
\begin{equation} \label{b3}
\tilde{P}^* (\vec{s}) = \prod_i \tilde{p}^* (s_i) = \prod_i \left(\frac{z}{z+s_i}\right)^z.
\end{equation}
Translating this back to the force distribution we get (\ref{a10}).

Further information is deduced by differentiation of the identity (\ref{b1}) with
respect to $\lambda $ giving
\begin{equation} \label{b4}
\int D q_i \,\frac{\sum_\alpha q^\alpha_i b_\alpha}
{(1+ \lambda \sum_\alpha q^\alpha_i b_\alpha)^{z+1}} 
= \frac{1}{z} \,\left(\prod_\alpha \, \frac{1}{1 + \lambda b_\alpha}\right) \sum_{\alpha'} \,
\frac{b_{\alpha'}}{1 + \lambda b_{\alpha'}}.
\end{equation}
This allows us to construct another invariant
\begin{equation} \label{b5}
\tilde{P}^*_1 (\vec{s}) = \tilde{P}^* (\vec{s}) \, \sum_i \,\frac{s_i}{ 1+ \lambda s_i}.
\end{equation}
Inserting this expression into the recursion relation (\ref{a8}), we encounter the integral 
(\ref{b1}) for all sites, except for site $i$ where we have to perform
the integral of the left hand side of (\ref{b4}). That produces for a set of 
surrounding sites factors with the power $z+1$ and the numerics is such that
adding them together yields again a sum as in (\ref{b5}). It is not surprising
that this additional invariant exists since we showed that (\ref{b2}) was invariant
for arbitrary $\lambda $ or scale of the forces. Thus the invariance of (\ref{b5}) is
nothing else than expressing this scale invariance of the recursion. In fact (\ref{b5})
follows directly from differentiating (\ref{b2}) with respect to $\lambda $.

It is tempting and indeed rewarding to continue to differentiate the identity 
(\ref{b1}). Differentiation of (\ref{b4}) with respect to $\lambda $ yields
\begin{equation} \label{b6}
\int D q_i \frac{(\sum_\alpha q^\alpha b_\alpha)^2}
{(1 + \lambda \sum_\alpha q^\alpha b_\alpha)^{z+2}} =  \frac{1}{z(z+1)} 
\left(\prod_\alpha \frac{1}{1 + \lambda b_\alpha}\, \right) \sum_{\alpha', \alpha''}
{b_{\alpha'} (1 + \delta_{\alpha',\alpha''}) b_{\alpha''} 
\over (1 + \lambda b_{\alpha'}) \,(1 + \lambda b_{\alpha''})}.
\end{equation} 
Now the distribution 
\begin{equation} \label{b7}
\tilde{P} (\vec{s}) = \tilde{P}^* (\vec{s})\,\sum_i \left(\frac{s_i}{z + s_i}\right)^2,
\end{equation} 
is not invariant since the right hand side of (\ref{b6}) produces terms with
powers $z+1$ of the denominator on two sites. We will show in  next sections that 
(\ref{b7}) leads to the ``slowest mode'' decaying towards to the stationary state.

The fact that the distribution (\ref{b2}) is stationary for arbitrary values of $\lambda$ leads to
a multitude of other stationary distributions by differentiation of (\ref{b2}) with 
respect to $\lambda$. The distribution (\ref{b5}) is an example. In the Appendix we comment
on the relevance of these stationary states. From this section the main message is:
\begin{itemize}
\item In the transformation from layer to layer, the total power of the factors
remains the same. This implies that we can select classes of perturbations which 
transform into themselves.
\item The sum of the coefficients of the terms is invariant under the transformation.
\end{itemize}

\section{The slowest mode}\label{mode}

Let us now investigate how a deviation from $\tilde{P}^*(\vec{s})$ will decay 
under recursion. We take an initial distribution of the type
\begin{equation} \label{g1}
\tilde{P} (\vec{s}) = \tilde{P}^* (\vec{s}) [\,1 + \Delta \tilde{P} (\vec{s})\,].
\end{equation}
In the spirit of the previous section we consider deviations of the form
\begin{equation} \label{g5}
\Delta \tilde{P} (\vec{s}) = \sum_n A_n \,\tilde{Q}_n (\vec{s}),
\end{equation}
where the $Q_n (\vec{s})$ are products of two factors.
\begin{equation} \label{g2}
\tilde{Q}_n (\vec{s}) = \sum_i \frac{s_i}{z + s_i} \, {s_{i+n} \over z + s_{i+n}}.
\end{equation}

It is clear from the previous Section that a $\Delta \tilde{P} (\vec{s})$ of the form (\ref{g5})
is after the transformation again a sum over the $\tilde{Q}_n (\vec{s})$
\begin{equation} \label{g6}
\Delta \tilde{P}' (\vec{s})  = \sum_n A'_n \,\tilde{Q}_n (\vec{s}).
\end{equation}
The transformed coefficients $A'_n$ are expressed in terms of the $A_n$ of 
the previous layer, using the formulae of the previous section,  as
\begin{equation} \label{g7}
A'_n = {1 \over z^2} \sum_{\alpha,\alpha'} A_{n+\alpha-\alpha'} + g_n \, A_0.
\end{equation}
On purpose we have chosen the same notation for the coefficients in the representation 
(\ref{g5}) and in Section \ref{corr} for the correlations in the forces. As one sees 
from comparing (\ref{g7}) with (\ref{z7}) and (\ref{z4}) the coefficients in both cases
obey exactly the same recursion relation. This means that we can take over the conclusions
of Section \ref{corr}. The first one is that we can construct a stationary state of the
form
\begin{equation} \label{g8}
A^*_n = B (1 + {1 \over z} \delta_{n,0}).
\end{equation}
The nature of this stationary state, having long--ranged correlations, will be
discussed in the Appendix. The other relevant conclusion 
concerns the case where only $A^{(0)}_0 \neq 0$. This is a minimal disturbance of the 
stationary state with only a modification of the single--site probability distribution and
no correlations between sites.\footnote{It is not a product distribution, but it can be
seen as the linearization of a product distribution.}  This is precisely the 
same initial condition as for the case of the second moment starting with an 
uncorrelated initial state. Thus the development 
of this deviation (\ref{g5}) is exactly the same as that of the second moment of an 
arbitrary deviation from the stationary state. The difference is that we now have the 
development of the full force distribution. Projecting this joint distribution onto a 
single-site distribution, by putting all $s_i=0$ except for one, one finds
\begin{equation} \label{new5}
\tilde{p}^{(l)}(s)=\tilde{p}^*(s) \left( 1+A^{(l)}_0 \frac{s^2}{(z+s)^2} \right).
\end{equation}
So on the level of the single--site distribution the shape of the deviation remains invariant,
but the amplitude decays as $A^{(l)}_0 \sim l^{(1-d)/2}$. This means that not only the 
second moment, but the whole single--site distribution relaxes towards the stationary state 
(\ref{a10}) with a characteristic shape of the single--site distribution given by 
(\ref{new5}). 

\section{Higher Modes} \label{modes}

In this section we show that higher moments lead to slower modes. Let us start by looking
at the third moment of the forces. The linearity of the force relation (\ref{a2}) 
guarantees that the third moments transform as a closed set 
\begin{equation} \label{h1}
\langle \, f_j f_{j+n} f_{j+m}\, \rangle' = \sum_{\alpha,\alpha',\alpha''} \left( \int D 
\vec{q} \,q^\alpha_{j-\alpha} \, q^{\alpha'}_{j+n-\alpha'} \,q^{\alpha''}_{j+m-\alpha''}
\right) \,\langle \, f_{j-\alpha} \,f_{j+n-\alpha'} \, f_{j+m-\alpha''}\,\rangle.
\end{equation}
The $q$ integral is complicated but elementary, with several exceptional cases due 
the equality of the lower indices. The general trend is however clear. One gets a relation 
of the form
\begin{equation} \label{h2}
\langle \, f_j f_{j+n} f_{j+m}\, \rangle' =  {1 \over z^3} \sum_{\alpha,\alpha',\alpha''}  
\langle \, f_{j-\alpha} f_{j+n-\alpha'} f_{j+m-\alpha''} \, \rangle + 
{\rm correction \, terms}.
\end{equation}
The correction terms refer to the cases where the indices $j,j+n$ and $j+m$ are either equal
or nearest neighbors. 

Consider now the class of uncorrelated distributions which coincide with the stationary 
state (\ref{a10}) up to the second moment but start to deviate at the level of the third 
moment. The difference of the third order correlations with respect to the stationary 
state then only has a value for $n=m=0$. Such a disturbance gradually spreads over the 
$2(d-1)$ dimensional space of the indices $n$ and $m$. The decay rate is therefore as 
$l^{(1-d)}$ which is faster than the $l^{(1-d)/2}$ of the slowest mode. The correction 
terms in (\ref{h2}) mildly modify the spreading of the correlations described 
by the first term on the right hand side of (\ref{h2}). They do no change the power 
of the decay but change the amplitude in front of the power.

The same story holds for the construction of the higher modes. By further differentiation 
of relation (\ref{b6}) one derives the transformation law of the functions which
have 3 factors of the type $s_i /(z+s_i)$ extra over the stationary state. Such functions
do not modify the normalisation, the mean force and the second moment and therefore
couple to the decay of the third moment as we have described above. On the level of the
single--site distribution the mode takes the form
\begin{equation} \label{h3}
\tilde{p} (s) = \tilde{p}^* (s) \left[1 + B^{(l)}_{0,0} 
\left({s \over z + s} \right)^3 \right],
\end{equation}
with the $B^{(l)}_{0,0} \sim l^{(1-d)}$. 
\begin{figure}[h]
   \centering {\epsfxsize=7cm
   \epsffile{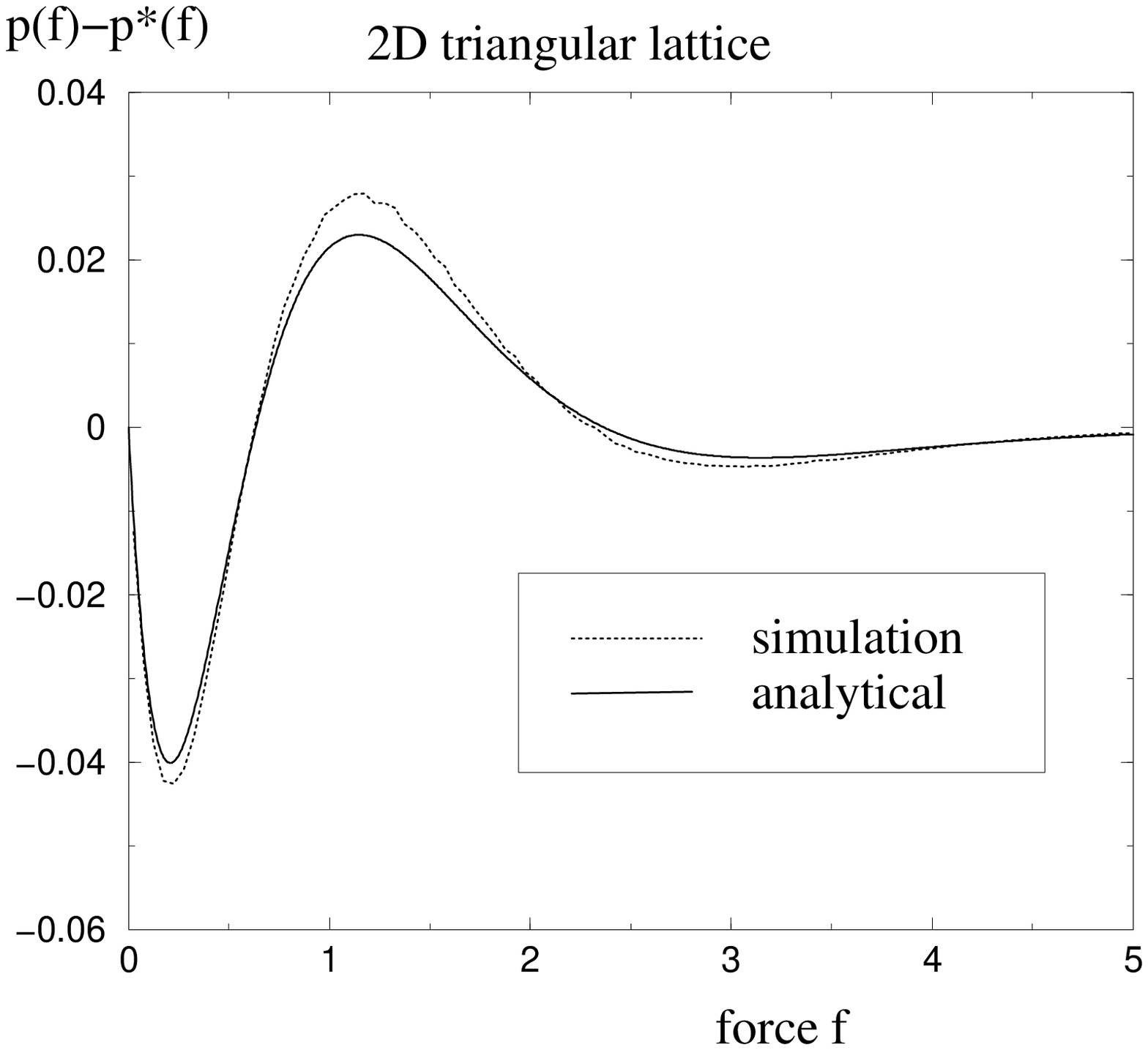} \epsfxsize=7cm \epsffile{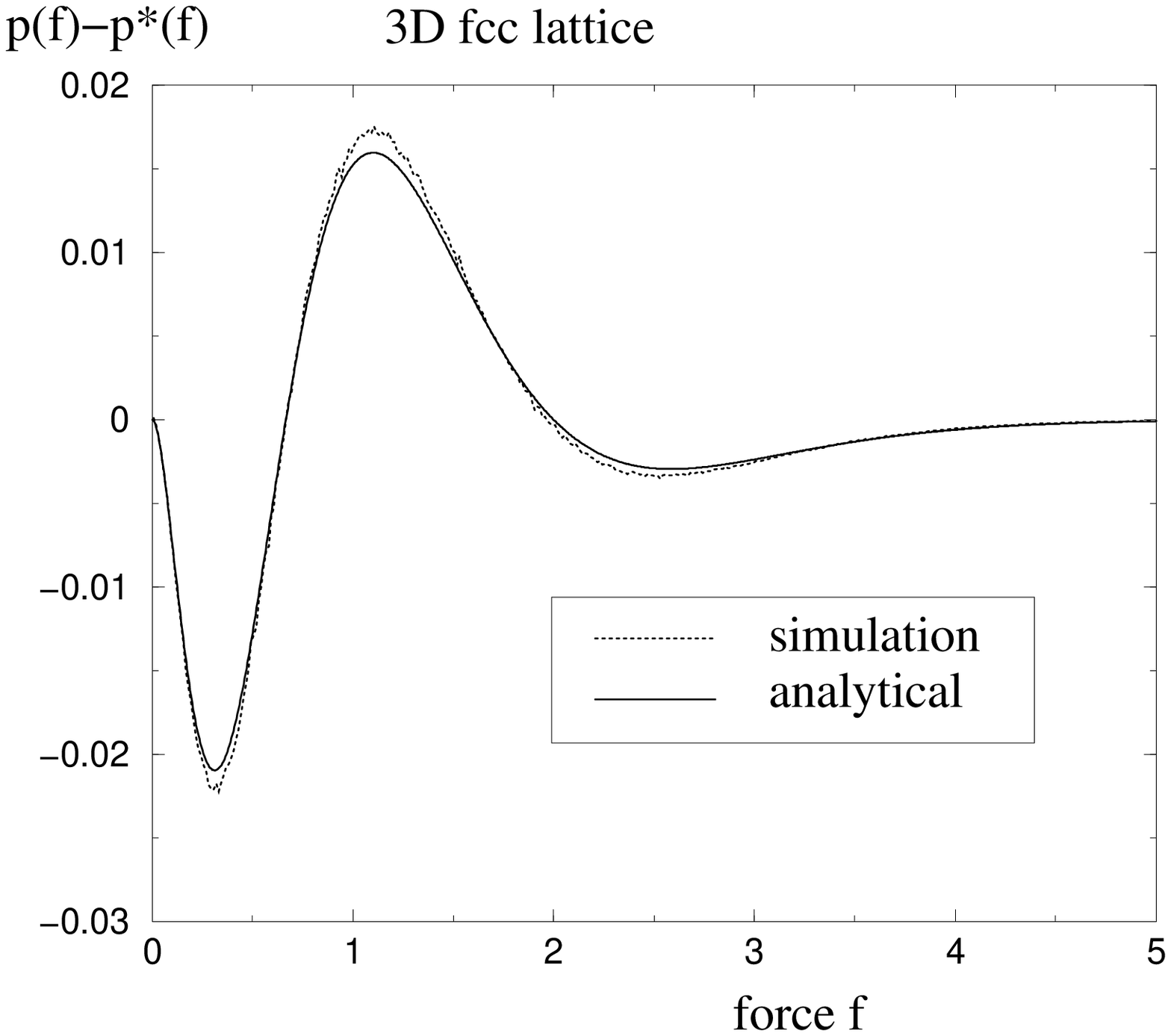}}
   \caption{The slowest mode for the triangular packing at $l=50$ and the fcc packing at $l=10$}
   \label{mds}
\end{figure}

The generalization to higher moments and faster modes is evident. On the level
of the single site distribution it simply means a change of the power of the last factor in
(\ref{h3}). The coefficient in front of a power $m$ decays as  $l^{(1-d)(m-1)/2}$.
 
The question that emerges is: can one observe these modes in the relaxation of an actual
simulation? This question is related to the problem of whether an arbitrary initial distribution
can be decomposed as a superposition of these modes. Already on the level of the 
single site distribution one observes that our modes do not form
a complete set in the mathematical sense. All modes start with a power $f^{(z-1)}$ in
the force distribution;
hence the modes form a poor basis for the small $f$ behavior. Nevertheless, the modes
that we have constructed do describe the asymptotic decay. We should realize that before
the asymptotic regime sets in, a fast process takes place. If we start out with a distribution
with a finite probability density for small forces, this probability will be supressed rapidly
in the region of small forces by the recursion relation, due to the available phase space.
For a small resultant force $f'_j$ to occur, all the components $q^\alpha_{j-\alpha}$ 
have to be small. Already after one recursion step, a finite probability density (in the 
single--site distribution) will be replaced by a probability density starting as $f^{z-1}$ 
(modulo logarithmic factors, which move to higher orders in the next recursion
steps). Thus after a few steps, the decay of the probability distribution can be well described
by the above constructed modes.

In order to test the mode picture we have performed numerical simulations on a 
2--dimensional triangular lattice and a 3--dimensional fcc lattice. We took the initial 
condition $f_i=1$ for each site $i$, which also means $\langle \, f_i^2 \, \rangle =1$. 
This initial distribution is a substantial deviation from the stationary state (\ref{a10}).
In Fig. \ref{mds} we have compared the observed single--site distributions with the 
analytical results. In the first figure, we show the single--site force distribution of the
$l=50$th layer in the triangular packing, after subtracting $p^* (f)$.  The analytical curve 
is the slowest mode (\ref{new5}), whose amplitude has been computed from the 
recursive scheme (\ref{g7}), with the initial value $A^{(0)}_0 =-1$, corresponding to the
initial second moment $\langle \, f_i^2 \, \rangle =1$. 
As the comparison does not involve any free parameters the correspondence is remarkable;
the more so since the deviations bear the signature of the next slowest mode (\ref{h3}).
In the second figure, the simulation data is shown for the fcc packing
with the same initial distribution and $l=10$. In line with the fact that the modes decay
faster in a higher dimension, the agreement is even more impressive.
  
\section{Discussion}

We have calculated the decay towards the stationary state (\ref{a10}) of initial
force distributions which have short--ranged force correlations and a sufficiently sharp
distribution of the total force (\ref{vn1}). Since we can fix the average force, the
relaxation occurs for the second and higher order force correlations.
The short--ranged correlations become longer ranged while their amplitudes diminish 
in a diffusion process in correlation space. This accounts for the slow algebraic relaxation 
of the second moments of the force. These moments also dictate the 
slowest modes of relaxation of the full distribution function. For arbitrary initial conditions
we find the asymptotic form (\ref{new5}) for the relaxation of the single--site distribution.
We have focussed on the triangular and the fcc packing as examples, but the results can be
easily extended to other packings and other connections between layers. The number
of connections  $z$ determines the form of the stationary state and the decay
modes. The dimension of the system detemines the relaxation powers of the
modes.

The stationary state (\ref{a10}) is not the only stationary state as can already be seen
from the fact that each initial distribution of the total force stays invariant. If we
exclude, however, long--ranged correlations in the initial state and restrict ourselves to
sharp distributions of the total force in the sense of (\ref{vn1}), the stationary state
can only marginally differ from (\ref{a10}) as is explained in the appendix.

It is not remarkable that the product stationary state (\ref{a10}) results also from a 
mean--field approximation. However, it is interesting that the single--site distributions 
(\ref{new5}) etc. are also found as modes of the mean--field approximation. 
Yet it would be misleading to conclude that the mean--field approximation is accurate. 
The mean--field modes
decay exponentially as a consequence of ignoring correlations in the system.
An initially uncorrelated state develops correlations which get longer in range and
smaller in amplitude. The long persistence of these correlations explains why the
mean--field approximation does not accurately describe the relaxation.

In our calculations we have relied heavily on the mathematical simplifications 
due to the uniform distribution of the $q$ variables. The slow approach to the 
stationary state is a robust property which holds for arbitrary $q$ distributions.
In fact it is easy to show that the decay of the second moments for an arbitrary $q$  
distribution is similar to what has been found in Section \ref{corr} (see also \cite{lew}). 
However, the identity  (\ref{b1}) only holds for the uniform distribution. So there is as yet
no general procedure to construct the explicit form of the stationary state and the 
slow modes. Coppersmith et al.  
\cite {cop} have found a countable set of $q$ distributions for which the stationary
state is uncorrelated, on the basis of a generalization of (\ref{b1}). For these distributions
the construction of the modes is completely similar to what we have done for the
uniform distribution. In general, the asymptotic distribution will exhibit correlations, 
albeit in higher moments than the second. The question how the correlations in the 
asymptotic state are related to the properties of the $q$ distribution remains 
intriguing, even if the correlations are of higher order and likely to be 
small in magnitude.

{\bf Acknowledgements}. The authors are indebted to Wim van Saarloos for posing the problem
and continued support, as well as to Martin van Hecke, Ellak Somfai and Martin
Howard for illuminating discussions.

\appendix
\section{Other Stationary States} \label{stat}

There are two ways to investigate the occurrence of a stationary state. We can do this on
the level of the moments or of the full probability distribution. As the moments are
simpler we consider first the second moment. We take the total force distribution 
to be sharp according to (\ref{vn1}). Imposing this condition on the stationary state
(\ref{z6}) gives for the coefficient $C$
\begin{equation} \label{A2}
C = {1 + c/N \over 1 + 1/zN} \simeq 1 + { c-1/z \over N}.
\end{equation}
So indeed the stationary values (\ref{z6}) approach those of (\ref{a10}) in the limit
$N \rightarrow \infty$. 

Similar arguments can be given for the stationary state (\ref{b5}) following from the scale 
invariance of the transformation.  This state does not 
correspond to a good distribution as its norm is zero. This problem can be avoided by 
adding $\tilde{P}^*(\vec{s})$ and allowing for a different scale, i.e.
\begin{equation} \label{A5}
\tilde{P}^*_1 (\vec{s}) = \tilde{P}^* (\lambda \vec{s}) \left( 1 + A \sum_i \frac{s_i}
{1+\lambda s_i} \right).
\end{equation}
Unlike the slowest mode, this particular addition to $\tilde{P}^* (\vec{s})$ does affect the 
average force. Rescaling the forces such that $\langle \, f_i \, \rangle =1$, requires 
$\lambda z = 1-A$. The second moments of (\ref{A5}) are
\begin{equation} \label{A7}
\langle \, f_i f_{i+n} \, \rangle = (1 - A^2) \, \frac{z+\delta_{0,n}}{z}.
\end{equation}
Thus the distribution is sharp in the sense of Eq. (\ref{vn1}) if
\begin{equation} \label{A6}
A ^2 \simeq (1/z-c)/N,
\end{equation}
which means that $A$ is ${\cal{O}} (1/ \sqrt{N})$ (or 0 for $c=1/z$ corresponding to the 
stationary state (\ref{a10})). Thus the stationary state (\ref{A5}) becomes equal to the
stationary state (\ref{a10}) in the thermodynamic limit. 
The long--ranged correlations are indeed of the order $1/N$.   

The same conclusion can be drawn with respect to the stationary state corresponding to
(\ref{g8}) which reads
\begin{equation} \label{A3}
\tilde{P}^*_2 (\vec{s}) = \tilde{P}^* (\vec{s}) \left( 1 + B \sum_n Q_n (\vec{s}) +
{B \over z} Q_0 (\vec{s}) \right).
\end{equation}
The allowed values for $B$ follow again from computing $\langle F^2 \rangle$ for the 
distribution (\ref{A3}) and equating this with the expression (\ref{vn1})
\begin{equation} \label{A4}
(1+ 2 B/z^2) (N^2 + N/z) = N^2 + c N \quad \quad {\rm or} \quad \quad B 
\simeq z^2(c-1/z)/2N.
\end{equation}
So $B$ turns out to be ${\cal{O}} (1/N)$ showing that the stationary state
implied by (\ref{g8}) is indeed only a marginal deviation from the stationary state 
(\ref{a10}). It has long--ranged correlations of the order of $1/N$. Similar 
arguments can be given for the other stationary states following from the scale 
invariance of the transformation. For finite systems such states are different, but in
the thermodynamic limit they all merge in the stationary state (\ref{a10}) if we 
insist on the sharpness of the total force distribution. 

\end{document}